\documentclass[aps,prb,10pt,twocolumn,superscriptaddress,
  showpacs,amsmath,amssymb,floatfix]{revtex4-1}

\usepackage[dvipdfmx]{graphicx}
\usepackage{bm}
\usepackage{braket}

\newcommand{\pd}{\partial}

\renewcommand{\r}{\bm{r}} 
\renewcommand{\k}{\bm{k}}

\renewcommand{\c}{\mathrm{c}}
\renewcommand{\i}{\mathrm{i}}

\newcommand{\tix}{\widetilde{x}}
\newcommand{\tiy}{\widetilde{y}} 

\newcommand{\De}{\Delta}

\newcommand{\x}{\mathrm{x}}
\newcommand{\y}{\mathrm{y}}
\newcommand{\z}{\mathrm{z}}

\newcommand{\FS}{\mathrm{FS}}

\newcommand{\px}{p_{\x}}
\newcommand{\py}{p_{\y}}
\newcommand{\dzx}{d_{\z\x}}
\newcommand{\dyz}{d_{\y\z}}

\newcommand{\tbeta}{\widetilde{\beta}} 
\newcommand{\trx}{\widetilde{\x}}
\newcommand{\try}{\widetilde{\y}}  

\newcommand{\dv}{\bm{d}}

\newcommand{\etaa}{\eta_{\mathrm{\,a}}}
\newcommand{\etab}{\eta_{\mathrm{\,b}}}
\newcommand{\phia}{\phi_{\mathrm{a}}}
\newcommand{\phib}{\phi_{\mathrm{b}}}

\begin{document}

\title{Chiral superconductivity in nematic states}

\author{Shuhei Takamatsu}
\email[]{}
\affiliation{Graduate School of Science and Technology, Niigata University,
  Niigata, 950-2181, Japan.}

\author{Youichi Yanase}
\affiliation{Department of Physics, Niigata University,
  Niigata, 950-2181, Japan.}

\date{\today}

\begin{abstract}
We investigate chiral superconductivity which occurs in the electronic nematic state. 
A vortex state in a $c$-axis magnetic field is studied on the basis of the two-component Ginzburg-Landau 
model for nematic-chiral superconductors. It is shown that various vortex lattice structures are 
stabilized by nontrivial cooperation of nematicity and chirality in superconductors. 
In particular, the vortex lattice structural transition occurs when a square anisotropy parameter $\nu$ 
is positive (negative) and the nematicity is induced along the [110] axis ([100] axis). 
We discuss nematic-chiral superconductivity in URu$_2$Si$_2$, Sr$_2$RuO$_4$, and UPt$_3$. 
An experimental test for the examination of nematic order and chiral superconductivity is proposed. 
\end{abstract}


\maketitle

\section{Introduction}
Recent studies on the strongly correlated electron systems have explored
the chiral superconductivity with broken time-reversal symmetry. 
A chiral ($\px \pm \i \py$)-wave superconductivity analogous to
the $^3$He A phase has been established in
Sr$_2$RuO$_4$ \cite{Mackenzie2003,Maeno2012}, and
a chiral ($\dzx \pm \i \dyz$)-wave superconductivity in
URu$_2$Si$_2$ has been identified by recent experimental studies
\cite{PRL_99_116402_Kasahara,PRL_100_017004_Yano,JPSJ_Kawasaki,Kapitulnik_URu2Si2,Yamashita_Nature_Phys}. 
Spontaneous time-reversal symmetry breaking characteristic of chiral superconductors has been observed 
in both compounds \cite{JPSJ_Kawasaki,Kapitulnik_URu2Si2,Luke_Sr2RuO4,Xia_Sr2RuO4}. 
Furthermore, a recent polar Kerr rotation measurement detected 
broken time-reversal symmetry in UPt$_3$\cite{Kapitulnik_UPt3},
and thus a chiral superconducting (SC) state which belongs to the $E_{\rm 1u}$, $E_{\rm 2u}$, 
or $E_{\rm 1g}$ representation is suggested.

Interestingly, chiral superconductivity coexists with a nematic order
in at least some of these superconductors. 
Nematic states in itinerant electron systems analogous
to classical liquid crystals have been one of the highlights
in recent condensed-matter physics \cite{Fradkin}. 
A "quantum nematic liquid crystal" accompanying
spontaneous rotation symmetry breaking has been studied 
in some strongly correlated electron systems. 
For instance, a nematic order arising from the fluctuating stripe order
has been proposed for cuprate superconductors \cite{Kivelson_2003}, and a nematic state
in a bilayer ruthenate Sr$_3$Ru$_2$O$_7$ has been investigated extensively \cite{Mackenzie_Sr3Ru2O7}. 
Furthermore, a nematic fluctuation in Fe-based superconductors
has been investigated \cite{Fernandes,Goto} 
and identified as a possible glue of Cooper pairs \cite{Kontani,Yanagi}.

There is accumulating evidence for a nematic order
in the so-called hidden-ordered state of URu$_2$Si$_2$ \cite{Mydosh_RMP}. 
The magnetic torque \cite{Okazaki},
cyclotron resonance \cite{Tonegawa_cycrotron}, 
NMR \cite{Kambe}, and x-ray scattering \cite{Tonegawa_Xray} measurements 
uncovered a broken fourfold rotation symmetry below
the hidden-order temperature $T < T_{\rm HO}$, 
although another NMR measurement did not detect any broken symmetry \cite{Mito}. 
Since the SC transition is a continuous second-order phase transition, 
the nematic order has to coexist with the superconductivity below $T_{\c}$. 
Indeed, a signature of nematic order has been observed in the SC state \cite{JPSJ_79_084705_Okazaki}. 
A kink in the lower critical field $H_{\c 1}$ has been attributed to
a second SC transition due to the broken fourfold rotation symmetry. 
It is also known from early experimental results \cite{Adenwalla,Hayden} 
that the sixfold rotation symmetry in the hexagonal crystal lattice 
of UPt$_3$ is broken in the SC state.
The splitting of two transition temperatures and the existence of
the A phase at zero magnetic field has been attributed to
the effect of a nematicity \cite{Adv_Phys_43_113_Sauls,Joynt}. 
Although the origin of broken rotation symmetry is still unclear, 
the antiferromagnetic order \cite{Hayden,Aeppli} may cause the nematicity. 
What is clear is that a nematicity plays an essential role in
the multiple SC phases in UPt$_3$. 
Although a spontaneous nematic order does not occur in Sr$_2$RuO$_4$, 
a nematicity is artificially induced by uniaxial pressure. 
Indeed, an enhancement of the transition temperature due to
the uniaxial pressure has been observed
in Sr$_2$RuO$_4$ \cite{Science_344_18_Hicks}. 
Thus, all the chiral superconductors known up to now coexist with the nematicity.

It is interesting to study the vortex state of these nematic-chiral superconductors
because of the following two reasons. 
First, an effect of nematic order competes with the gradient mixing in order parameters 
due to a magnetic field. 
Although the nematicity favors a nonchiral state,
such as the ($\px \pm \py$)-wave state or ($\dzx \pm \dyz$)-wave state, 
the magnetic field along the $c$ axis stabilizes the chiral state, 
such as the ($\px \pm \i \py$)-wave state or ($\dzx \pm \i \dyz$)-wave state, through the gradient mixing. 
Therefore, the order parameter in nematic-chiral superconductors is nontrivial. 
Second, various vortex lattice structures are stabilized
in the chiral superconductors 
because of the gradient coupling of two-component order parameters 
\cite{Agterberg_PRL_80_5184,Agterberg_PRB_58_14484,Kita_PRL_83_1846}. 
Thus, it is expected that rich vortex lattice phases appear in nematic-chiral superconductors. 
From the other perspective, the vortex state in chiral superconductors is sensitive to the nematicity 
in the underlying electronic state.
Therefore, the vortex lattice structure will be a sensitive probe for detecting the nematic order.

In this paper we investigate the vortex state of nematic-chiral superconductors
on the basis of the Ginzburg-Landau (GL) theory.
Signatures of the nematic order in the chiral SC state are clarified. 
The organization of this paper is as follows. In Sec.~II,
we introduce the GL model and describe a variational method by which we study the vortex state.
In Sec.~III, we show the phase diagram for the order parameter and vortex lattice structure. 
Experimental results on URu$_2$Si$_2$, Sr$_2$RuO$_4$,
and UPt$_3$ are discussed and future experimental studies are proposed in Sec.~IV.

\section{Ginzburg-Landau Theory}
\subsection{Ginzburg-Landau model}
In this section, we construct a two-component Ginzburg-Landau (GL) model
for chiral superconductors in the nematic state.  
Essential variables are two-component order parameters, $(\etaa (\r),\etab (\r))$,
by which the pairing function is described as
$\De(\r,\k) = \etaa (\r) \phia (\k) + \etab (\r) \phib (\k)$. 
Here, $\phia (\k)$ and $\phib (\k)$ stand for pairing functions in the momentum space 
which belong to a two-dimensional representation of the crystal point group \cite{Rev_Mod_Phys_63_239}. 
For instance, the chiral ($\px \pm \i \py$)-wave SC state in Sr$_2$RuO$_4$ belongs to 
the $E_{\rm u}$ representation of the $D_{\rm 4h}$ point group \cite{Sigrist_review_1999}, 
while the chiral ($\dzx \pm \i \dyz$)-wave state in URu$_2$Si$_2$
belongs to the $E_{\rm g}$ representation. 
The chiral $f$-wave superconductivity, which belongs to the $E_{\rm 2u}$ representation of
the $D_{\rm 6h}$ point group, has been suggested for a representative multicomponent 
superconductor UPt$_3$ \cite{Adv_Phys_43_113_Sauls}. 
On the other hand, a recent thermal conductivity measurement indicated
the $E_{\rm 1u}$ representation \cite{Machida_Izawa}. 
All of these chiral SC states are represented by two-component order parameters
$(\etaa (\r),\etab (\r))$ as above, 
although the pairing functions $\phia (\k)$ and $\phib (\k)$ depend on materials.

First, we discuss a GL model in the absence of the nematicity. 
Assuming a weak nematicity, we later add a quadratic symmetry-breaking term. 
The symmetric part of the GL free-energy density has been obtained as 
\begin{align}
  F_{0} &= \alpha (|\etaa |^2+|\etab |^2 )
  +\frac{\beta_1}{2} (|\etaa |^2+|\etab |^2 )^2 \nonumber\\
  &\quad +\frac{\beta_2}{2}(\etaa \etab^* - \c.\c.)^2 
  +\beta_3 |\etaa |^2 |\etab |^2  \nonumber\\
  &\quad +\kappa_1 ( |D'_{\x}\etaa |^2 + |D'_{\y}\etab |^2 ) \nonumber\\
  &\quad +\kappa_2 ( |D'_{\x}\etab |^2 + |D'_{\y}\etaa |^2 ) \nonumber\\
  &\quad +\kappa_3 \bigl[(D'_{\x}\etaa )(D'_{\y}\etab )^* + \c.\c.\bigr] \nonumber\\
  &\quad +\kappa_4 \bigl[(D'_{\x}\etab )(D'_{\y}\etaa )^* + \c.\c.\bigr] 
  \label{eq:f0}
\end{align}
for chiral superconductors in the type-II limit \cite{Rev_Mod_Phys_63_239}. 
Following the conventional notation, we denote
$\alpha = \alpha_{0} (T/T^0_{\c} - 1)$, with $T^0_{\c}$
being the transition temperature at zero magnetic field, and covariant derivatives 
$D'_{j}= -\i \pd_{j}+(2 \pi/\varPhi_{0})A_{j}$, with $\varPhi_{0}=hc/2|e|$.
We omit gradient terms containing $z$ derivatives
because we focus on the vortex state in magnetic fields applied along the $c$ axis. 

In the weak-coupling BCS theory, parameters are given by the following relations: 
$\beta_2 / \beta_1 = \braket{\phia^2 \phib^2}_{\FS} / \braket{\phia^4}_{\FS}$,
$\beta_3 / \beta_1 = 3\beta_2 / \beta_1 - 1$,
$\kappa_2/\kappa_1 = \braket{\phia^2 v^2_{\y}}_{\FS} / \braket{\phia^2 v^2_{\x}}_{\FS}$,
and 
$\kappa_3/\kappa_1 = \kappa_4/\kappa_1 = 
\braket{\phia \phib v_{\x}v_{\y}}_{\FS} / \braket{\phia^2 v^2_{\x}}_{\FS}$,
where $v_{\x}$ and $v_{\y}$ are Fermi velocities in the $ab$ plane
and brackets $\braket{\cdots}_{\FS}$ denote an average over the Fermi surface. 
Despite three independent parameters, $\beta_2 / \beta_1$, $\kappa_2/\kappa_1$, and $\kappa_3/\kappa_1$, 
a single-parameter description of the GL model has been adopted in studies of 
chiral ($\px \pm \i \py$)-wave superconductivity 
\cite{Agterberg_PRL_80_5184,Agterberg_PRB_58_14484,Kita_PRL_83_1846}. 
As shown by Agterberg \cite{Agterberg_PRL_80_5184,Agterberg_PRB_58_14484},  
the coupling constants are represented by a single-parameter $\nu$ as \cite{comment1}
\begin{align}
  \beta_2/\beta_1 = \kappa_2/\kappa_1 = \kappa_3/\kappa_1 = (1+\nu)/(3-\nu)  
  \label{eq:nu}
\end{align}
when we assume pairing functions $(\phia (\k),\, \phib (\k))=(v_{\x}(\k),\, v_{\y}(\k))$ 
(Agterberg model). 
Adopting similar pairing functions for the chiral ($\dzx \pm \i \dyz$)-wave SC state, 
$(\phia(\k),\, \phib(\k)) = (v_{\x}(\k)f(k_{\z}),\, v_{\y}(\k)f(k_{\z}))$, 
with $f(k_{\z})$ being an odd function (extended Agterberg model), 
we approximately obtain parameters shown in Eq.~(\ref{eq:nu}). 
Strictly speaking, the single-parameter description breaks down 
since $\beta_2/\beta_1 \ne \kappa_2/\kappa_1 = \kappa_3/\kappa_1$. 
However, the deviation is negligible, $\beta_2/\beta_1 - \kappa_2/\kappa_1 \ll 1$,
for a smooth function $f(k_{\z})$. 
Thus, the single-parameter description is also applicable to
the chiral ($\dzx \pm \i \dyz$)-wave superconductors.
In crystals satisfying the $D_{\rm 4h}$ point group, 
the parameter $\nu$ indicates the square anisotropy in the Fermi surface,
because $\nu = 0$ for the cylindrical or spherical Fermi surface, while $\nu \ne 0$ otherwise. 
On the other hand, the chiral $E_{\rm 1u}$ state in the $D_{\rm 6h}$ point-group symmetry 
is similarly described as $(\phia (\k),\, \phib (\k)) = (v_{\x}(\k)f(k_{\z}),\, v_{\y}(\k)f(k_{\z}))$
with $f(k_{\z})$ being an even function. 
Then, we obtain $\beta_2/\beta_1 = \kappa_2/\kappa_1 = \kappa_3/\kappa_1 = 1/3$, and thus,
$\nu =0$ in Eq.~(\ref{eq:nu}) irrespective of the Fermi surface \cite{comment2}. 
Thus, we can rely on the single-parameter description of the GL model for 
these chiral $p$-wave, $d$-wave, and $f$-wave superconductivities
when the Fermi velocity and pairing functions have a smooth momentum dependence. 
Adopting the single-parameter description, we investigate the vortex state in chiral superconductors. 
On the other hand, Eq.~(\ref{eq:nu}) seriously breaks down in the chiral $E_{\rm 2u}$ state 
in the $D_{\rm 6h}$ point-group symmetry. Thus, the chiral $E_{\rm 2u}$ state is beyond the scope of our study, 
but we will briefly discuss it in Sec.~IV. 

\begin{figure}[htbp]
  \centering
  \includegraphics[width=5.25cm,origin=c,keepaspectratio,clip]{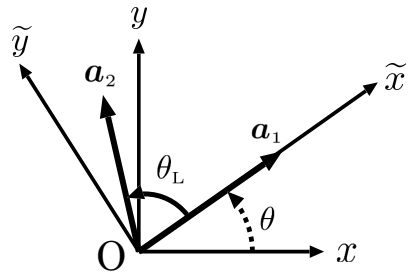}  
  \caption{Primitive vectors of the vortex lattice $\bm{a}_1$ and $\bm{a}_2$.
    Two-dimensional coordinates $(x,y)$ and $(\tix,\tiy)$ and
    angles $\theta_{\rm L}$ and $\theta$ are illustrated.} 
  \label{fig:unit_cell}
\end{figure}

We rewrite the GL model using order parameters in the chirality basis, 
$\eta_{1} = \gamma (\etaa -\i \etab ) /\sqrt{2}$ and
$\eta_{2} = \gamma (\etaa +\i \etab ) /\sqrt{2}$ with, 
$\gamma = (\tbeta_{1} / \alpha_0)^{1/2}$ and 
$\tbeta_{1} = \beta_{1}-\beta_{2} + \beta_{3}/2 = 2 \beta_{1}/(3-\nu)$. 
With the two-dimensional coordinate rotated through an angle $\theta$ around the $z$ axis (see Fig.~1), 
the dimensionless free-energy density is given by 
\begin{align}
  \hspace{-3mm}
  f_0 & = \frac{T - T^{0}_{\c}}{T^0_{\c}} ( |\eta_{1}|^2+|\eta_{2}|^2 )
  +\frac{1}{2} ( |\eta_{1}|^4+|\eta_{2}|^4 ) \nonumber\\
  &\quad + 2|\eta_{1}|^2|\eta_{2}|^2 - \frac{\nu}{2}[(\eta_{1}\eta^*_{2})^2 + \c.\c.]
  \nonumber\\
  &\quad + |D_{\trx}\eta_{1}|^2 + |D_{\try}\eta_{1}|^2 + |D_{\trx}\eta_{2}|^2 + |D_{\try}\eta_{2}|^2
  \nonumber\\
  &\quad + \frac{1}{2} \{ (e^{2\i\theta}-\nu e^{-2\i\theta}) \nonumber\\ 
  &\quad \times 
      [(D_{\trx}\eta_{1})(D_{\trx}\eta_{2})^* - (D_{\try}\eta_{1})(D_{\try}\eta_{2})^* ] + \c.\c. \} \nonumber\\
  &\quad + \frac{1}{2} \{ \i (e^{2\i\theta}+\nu e^{-2\i\theta}) \nonumber\\
  &\quad \times
      [(D_{\trx}\eta_{1})(D_{\try}\eta_{2})^* + (D_{\try}\eta_{1})(D_{\trx}\eta_{2})^* ] + \c.\c. \}, 
  \label{eq:f0_dim_less}
\end{align}
where the unit of energy, length, and magnetic field are 
$\alpha^2_{0} / \tbeta_{1}$, $\xi=[(\kappa_1+\kappa_2)/2\alpha_{0}]^{1/2}$,
and $\varPhi_{0}/2 \pi \xi^2$, respectively. 
The covariant derivatives are denoted as $D_{\, \widetilde j} =-\i \pd_{\, \widetilde j} + A_{\, \widetilde j}$. 
We focus on the magnetic field along the $c$ axis, and choose the vector potential 
$\bm{A} = - \tiy H \bm{e}_{\trx}$.
In the reasonable parameter range $|\nu| \leq 1$, 
a chiral SC state with $(\eta_{1},\eta_{2}) \propto (1,0)$ or $(\eta_{1}, \eta_{2}) \propto (0,1)$ 
is stable at zero magnetic field.

A weak nematicity leading to the violation of fourfold or sixfold rotational symmetry is taken into account 
by adding the symmetry-breaking term in the quadratic form 
\begin{align}
  f_{\rm 2h}  = g(\i \eta_{1}\eta^*_{2} + \c.\c.)
  = g \gamma^2 (\etaa \etab^* + \c.\c.) 
  \label{eq:f2h}
\end{align}
to $f_0$. 
Thus, the total free-energy density is given by 
\begin{equation}
  f = f_{0} + f_{\rm 2h}.
  \label{eq:f_tot}
\end{equation}
The coupling constant $g$ represents a lifting of degeneracy between two orbital pairing functions 
$\phia (\k)$ and  $\phib (\k)$ due to the nematic order. 
As a consequence of the symmetry-breaking term, double SC transitions occur at zero magnetic field. 
In the high-temperature phase near the SC transition temperature, $T_{\c 2} < T < T_{\c}$, 
a nonchiral state where $(\eta_{1},\eta_{2}) \propto (1,\pm \i)$ is stabilized,
while a chiral state is stabilized below $T_{\c 2}$. 
The time-reversal symmetry is spontaneously broken in the low-temperature phase. 
The double SC transition has been observed in UPt$_3$ \cite{Adenwalla,Hayden,Adv_Phys_43_113_Sauls,Joynt}, 
and a signature of double SC transition has been reported in URu$_2$Si$_2$ \cite{JPSJ_79_084705_Okazaki}.

In Eq.~(\ref{eq:f2h}) we assume a nematicity along the [110] direction to be consistent with  
experimental observations in URu$_2$Si$_2$ \cite{Okazaki, Tonegawa_cycrotron, Kambe, Tonegawa_Xray}. 
The nematicity along the [100] axis is also investigated on the basis of our model. 
When we change the coordinate $\theta \rightarrow \theta + \pi/4$ and take the phase factor 
$(\eta_1, \eta_2) \rightarrow (\eta_1 e^{-\i \pi/4}, \eta_2 e^{\i \pi/4})$,
the $f_0$ term is almost invariant except for the sign reversal $\nu \rightarrow - \nu$.  
Then, the symmetry-breaking term changes to 
\begin{align}
  f^{'}_{\rm 2h} = g \left(\eta_{1}\eta^*_{2} + \c.\c. \right)
  = g \gamma^2 \left(|\etaa|^2 - |\etab|^2\right), 
  \label{eq:f2h'}
\end{align}
which is nothing but the symmetry-breaking term due to the nematicity along the [100] axis. 
A symmetry-breaking term of this form has been adopted for studies on multiple SC phases in 
UPt$_3$~\cite{Adv_Phys_43_113_Sauls,Joynt}. 
We change the sign of $\nu$ instead of considering the symmetry-breaking term in Eq.~(\ref{eq:f2h'}) 
for studies of nematicity along the [100] axis.

\subsection{Variational method}
We investigate the order parameters and vortex lattice structure using the variational method. 
We assume variational wave functions of Cooper pairs so that the solution of
the linearized GL equation is reproduced. 
First, we solve the linearized GL equation using the Landau-level expansion. 
Differentiating the quadratic terms in the $f_0$ term [Eq.~(\ref{eq:f0_dim_less})] 
with respect to $\eta_{1}$ and $\eta_{2}$, we obtain the linearized GL equation 
in the absence of the symmetry-breaking term, 
\begin{align}
  \lambda
  \left( \begin{array}{c}
    \eta_{1} \\
    \eta_{2} 
  \end{array} \right) 
  &=
  \frac{1}{l^2_{c}} 
  \left( \begin{array}{cc}
    1+2\Pi_{+}\Pi_{-}   &   e^{-2\i \theta} \Pi^2_{-} - \nu e^{2\i \theta} \Pi^2_{+} \\
    e^{2\i \theta}\Pi^2_{+} - \nu e^{-2\i \theta} \Pi^2_{-}   &   1+2\Pi_{+}\Pi_{-}
  \end{array} \right)
  \nonumber \\
  &\quad \times
  \left( \begin{array}{c}
    \eta_{1} \\
    \eta_{2} 
  \end{array} \right),
  \label{eq:GLeq}
\end{align}
where $l_{\c} = 1/\sqrt{H}$ 
and $\Pi_{\pm} = - (l_{\c}/\sqrt{2}) (D_{\trx} \pm \i D_{\try})$. 
The solution for the minimum eigenvalue $\lambda = \lambda_{\rm min}$  
is represented as
\begin{equation}
  \left( \begin{array}{c}
    \psi_{1+}(\r) \\
    \psi_{2+}(\r)
  \end{array} \right)
  =
  \sum_{n\geq 0}
  \left( \begin{array}{c}
    a_{4n}(\theta)  \varphi_{4n}(\r,\rho,\sigma) \\
    a_{4n+2}(\theta) \varphi_{4n+2}(\r,\rho,\sigma)
  \end{array} \right), 
  \label{eq:chirality+}
\end{equation}
where $\varphi_{n}(\r,\rho,\sigma)$ denotes the $n$th Landau-level wave function. 
The leading term is the lowest Landau level of the positive chirality component, and thus, 
$\left(\psi_{1+}(\r),\psi_{2+}(\r)\right)  \simeq \left(a_{0}(\theta)  \varphi_{0}(\r,\rho,\sigma), 0\right)$. 
We obtain another solution for the pairing state with a dominantly negative chirality, 
\begin{equation}
  \left( \begin{array}{c}
    \psi_{1-}(\r) \\
    \psi_{2-}(\r) 
  \end{array} \right)
  =
  \sum_{n\geq 0}
  \left( \begin{array}{c}
    b_{4n+2}(\theta) \varphi_{4n+2}(\r,\rho,\sigma) \\
    b_{4n}(\theta)  \varphi_{4n}(\r,\rho,\sigma)
  \end{array} \right), 
  \label{eq:chirality-}
\end{equation} 
where the leading term is the lowest Landau level of the negative chirality component, 
$\left(\psi_{1-}(\r), \psi_{2-}(\r)\right) \simeq \left(0,b_{0}(\theta) \varphi_{0}(\r,\rho,\sigma)\right)$. 
Coefficients $a_n(\theta)$ and $b_n(\theta)$ are numerically determined. 
We assume variational wave functions consisting of a linear combination 
of the two solutions: 
\begin{equation}
  \left(
  \begin{array}{c}
    \eta_{1}(\r) \\
    \eta_{2}(\r)
  \end{array}
  \right)
  =
  C_{+} \left(
  \begin{array}{c}
    \psi_{1+}(\r) \\
    \psi_{2+}(\r) 
  \end{array}
  \right)
  +
  C_{-} \left(
  \begin{array}{c}
    \psi_{1-}(\r) \\
    \psi_{2-}(\r) 
  \end{array}
  \right),
  \label{eq:Delta}
\end{equation}
where $|C_{+}|$ ($|C_{-}|$) represents the weight of Cooper pairs having dominantly positive (negative) 
chirality. 
The variational wave function is justified near the transition temperature and
for a small symmetry-breaking term $g$, although the reconstruction of higher Landau levels affects
the vortex lattice structure at low temperatures \cite{Kita_PRL_83_1846}.  
Our main result is concerned with the vortex lattice structural transition near $T_{\c}$, 
and thus, the variational wave function is appropriate.

In order to study the vortex lattice structure,
we adopt a general form of the $n$th Landau-level wave functions 
\cite{SC_rev_1_207} 
\begin{align}
  \varphi_{n}(\r,\rho,\sigma) 
  &= \frac{1}{\sqrt{2^n \pi^{1/2} n!}}
       \sum_{m} c_{m} e^{ 2\pi \i (m-1/2)\tix/a } \nonumber \\ 
  &\quad \times H_{n} \left( \frac{\tiy - y_m}{l_{\c}} \right) e^{-(\tiy - y_m)^{2} / 2 l^2_{\c}},
  \label{eq:n-th_Landau_level}
\end{align}
where $c_{m} = e^{\i \pi m(\rho+1-m\rho)}$, $y_{m} = l_{\c}\sqrt{2\pi\sigma} (m-1/2)$, and
$H_{n}(y)$ is the Hermit polynomials. 
The vortex lattice structure is determined by the variables $\rho=(b/a)\cos\theta_{\rm L}$ 
and $\sigma=(b/a)\sin\theta_{\rm L}$. 
Primitive vectors are 
$\bm{a}_1 = a \bm{e}_{\trx}$ and 
$\bm{a}_2 = b (\cos\theta_{\rm L}\bm{e}_{\trx} + \sin\theta_{\rm L}\bm{e}_{\try})
= a (\rho\bm{e}_{\trx} + \sigma\bm{e}_{\try})$ (see Fig.~1). 
The area of the unit cell is $|\bm{a}_1 \times \bm{a}_2| = ab \sin\theta_{\rm L} = 2 \pi l^2_{\c}$. 
The rectangular and centered rectangular lattices are formed for $\rho = 0$ and $\rho = 1/2$, respectively. 
The square and triangular lattices are special cases of them; $(\rho, \sigma) = (0,1)$ or 
$(\rho, \sigma) = (1/2,1/2)$ in the square lattice, and
$(\rho, \sigma) = (1/2,\sqrt{3}/2)$ in the triangular lattice.  
The angle of a primitive vector $\bm{a}_1$ from the original $x$ axis is $\theta$,
and it is dealt with as a variational parameter. 
Thus, variational parameters for the vortex lattice structure are $\rho$, $\sigma$, and $\theta$.

Substituting Eqs.~(\ref{eq:chirality+})-(\ref{eq:n-th_Landau_level})
into Eqs.~(\ref{eq:f0_dim_less}) and (\ref{eq:f2h}) and integrating over the unit cell, 
we obtain the free-energy density as 
\begin{equation}
  F(C_{+},C_{-},\rho,\sigma,\theta) = \braket{f}_{\rm uc}
  = \frac{1}{2\pi l^2_{\c}} \int_{\rm uc} f(\r) d^2 r.
  \label{eq:F_unit_cell}
\end{equation}
The brackets $\braket{\cdots}_{\rm uc}$ denote an average over the unit cell.
Optimizing variational parameters $(C_{+},C_{-},\rho,\sigma,\theta)$ to minimize the free-energy density, 
we determine the order parameters and vortex lattice structure.

\section{chiral superconducting states}
\subsection{Vortex lattice in the non-nematic state}
In this section, we investigate the vortex state in the absence of nematicity. 
Thus, we consider a tetragonal or hexagonal system and choose the parameter $g=0$. 
The effects of nematic order on the chiral SC state are studied in the following sections. 

\begin{figure}[htpb]
  \centering
  \includegraphics[width=7.0cm,origin=c,keepaspectratio,clip]{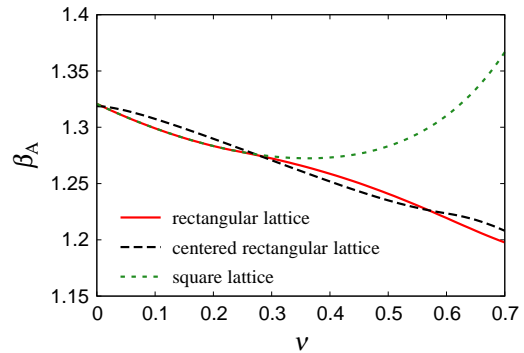}
  \caption{(Color online) 
    Abrikosov parameter $\beta_{\rm A}$ for the rectangular lattice ($\rho = 0$, red solid line)
    and centered rectangular lattice ($\rho = 1/2$, black long-dashed line) as a function of $\nu$.
    The variational parameter $\sigma$ is optimized to minimize $\beta_{\rm A}$. 
    The Abrikosov parameter for the square lattice [$(\rho, \sigma) = (0,1)$] 
    is shown by the green short-dashed line. 
    The other variational parameter $\theta$ is fixed to
    $\theta = 0$ ($\theta = \pi/4$) for $\nu > 0$ ($\nu < 0$), 
    for which the Abrikosov parameter is optimized when $|\nu| > 0.0114$. 
    When $|\nu| < 0.0114$, the optimized angle is $\theta = \pi/4$ ($\theta = 0$) 
    for $\nu > 0$ ($\nu < 0$), but the $\theta$ dependence of $\beta_{\rm A}$ is negligible.}
  \label{fig:betaA}
\end{figure}
\begin{figure}[htpb]
  \centering
  \includegraphics[width=7.0cm,origin=c,keepaspectratio,clip]{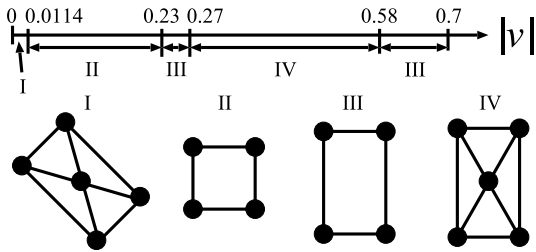}
  \caption{Schematic of the vortex lattice structure near the upper critical field. 
    We show the result for $\nu > 0$. 
    The vortex lattice is rotated $\pi/4$ for $\nu < 0$.}
  \label{fig:nu_dependence}
\end{figure}

First, we determine the vortex lattice structure near the upper critical field. 
Since the symmetry-breaking term $f_{\rm 2h}$ is absent, a solution of the linearized GL equation 
$(\eta_1, \eta_2) = C_+ (\psi_{1+}(\r), \psi_{2+}(\r))$ minimizes the free energy near the SC transition. 
Thus, $C_{-} =0$, and the vortex lattice structure is determined so as to minimize the Abrikosov parameter 
$\beta_{\rm A} = 2 \braket{f_{4}}_{\rm uc} /( \braket{|\psi_{1+}|^2}_{\rm uc} + \braket{|\psi_{2+}|^2}_{\rm uc})^2$,
where $f_{4}$ is the quartic term in Eq.~(\ref{eq:f0_dim_less}). 
Figure~\ref{fig:betaA} shows the Abrikosov parameter for various vortex lattice structures as a function 
of the anisotropy parameter $\nu$, and we illustrate the vortex lattice structure which minimizes 
the free energy in Fig.~\ref{fig:nu_dependence}. 
It is shown that the centered rectangular lattice, square lattice, rectangular lattice, and, again, 
centered rectangular lattice are stabilized with increasing $|\nu|$. 
These results are consistent with previous works which investigated the vortex lattice structure 
in the small-$|\nu|$ region \cite{Agterberg_PRL_80_5184, Agterberg_PRB_58_14484, Kita_PRL_83_1846}. 
Agterberg showed that the square lattice is stable unless the anisotropy parameter is extraordinary small, 
$|\nu| < 0.0114$ \cite{Agterberg_PRL_80_5184, Agterberg_PRB_58_14484}. 
The triangular lattice is formed at $\nu =0$ like in conventional superconductors, 
and it is deformed with increasing $|\nu|$.  
The parameter $\sigma$ decreases from $\sqrt{3}/2$ to $1/2$ when $|\nu|$ increases from 0 to 0.0114. 
The qualitatively same results have been obtained by Kita using a sophisticated calculation 
which takes into account higher Landau levels and screening current \cite{Kita_PRL_83_1846}. 
He also found that the rectangular lattice is stabilized for a large anisotropy parameter $|\nu|$. 
Our calculation reproduces these results and, furthermore, shows that the centered rectangular lattice 
and rectangular lattice are stabilized for $0.27 \leq |\nu| \leq 0.58$ and
$0.58 \leq |\nu| \leq 0.7$, respectively.
We do not consider a further large parameter $|\nu| > 0.7$ since the numerical convergence becomes worse.

\begin{figure}[htbp]
  \centering
  \includegraphics[width=7.25cm,origin=c,keepaspectratio,clip]{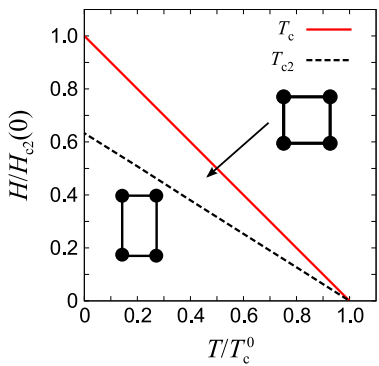}
  \caption{(Color online) Phase diagram of the vortex lattice as a function of
    the dimensionless magnetic field $H/H_{\c 2}(0)$ 
    and temperature $T/T^0_{\c}$ for $(\nu,g)=(0.15,0)$.
    The red solid line shows the SC transition temperature $T_{\c}(H)$, 
    while the black dashed line shows the second transition temperature $T_{\c 2}(H)$. 
    The vortex lattice structures are shown schematically. 
    The phase diagram is independent of the sign of $\nu$,
    while the vortex lattice for $\nu < 0$ is rotated $\pi/4$ from that for $\nu > 0$.}
  \label{fig:dia_nu=015_g=0}
\end{figure}

\begin{figure}[htbp]
  \centering
  \includegraphics[width=5.75cm,origin=c,keepaspectratio,clip]{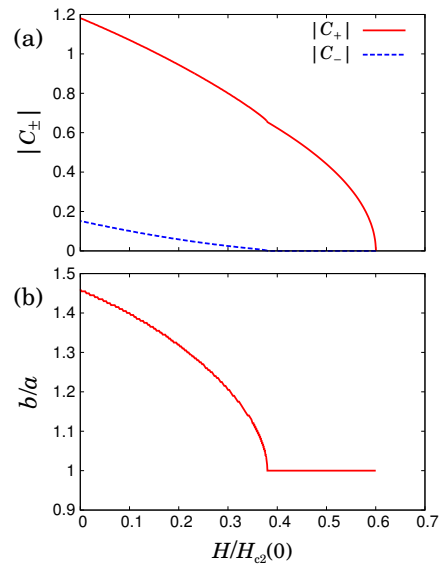}
  \caption{(Color online) Magnetic field dependence of (a) order parameters 
    $|C_{\pm}|$ and (b) a structural parameter $b/a$ $(=\sigma)$ at $T/T^0_{\c}=0.4$ for $(\nu,g)=(0.15,0)$.
    The other structural parameters are $\theta_{\rm L}=\pi/2$~($\rho=0$) and $\theta=0$.}
  \label{fig:nu=015_t=04}
\end{figure}

Next, we show the phase diagram against magnetic fields and temperatures ($H$-$T$ phase diagram) 
for two anisotropy parameters $\nu =0.15$ and $\nu=0.3$. 
For $\nu =0.15$, the square vortex lattice is stable near the transition temperature $T=T_{\c}(H)$,
as illustrated in Fig.~\ref{fig:nu_dependence}. 
Figure~\ref{fig:dia_nu=015_g=0} shows that the structural transition occurs at a moderate temperature 
$T=T_{\c 2}(H)$, and the rectangular vortex lattice is stabilized below $T_{\c 2}(H)$. 
Figure \ref{fig:nu=015_t=04} shows the magnetic field dependences of variational parameters $|C_{\pm}|$ 
and the "nematicity of the vortex lattice" $b/a$ at $T/T^0_{\c} = 0.4$. 
It is shown that the emergence of nematicity  $b/a-1$ coincides with $C_{-}$, although 
$C_{-}=0$ and $b/a=1$ in the high magnetic field region, $H/H_{\rm c2}(0)>0.38$. 
Thus, the structural transition in the vortex lattice accompanies the mixing of chirality in the 
order parameter. 

\begin{figure}[htbp]
  \centering
  \includegraphics[width=7.25cm,origin=c,keepaspectratio,clip]{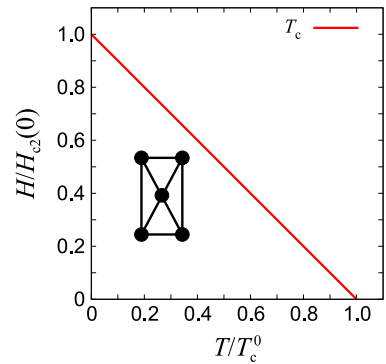}
  \caption{(Color online) Phase diagram of the vortex lattice for $(\nu,g)=(0.3,0)$.
  As shown schematically, the centered rectangular lattice is
  stabilized in the whole SC state. 
  }
  \label{fig:dia_nu=030_g=0}
\end{figure}

On the other hand, the vortex lattice structural transition does not occur for $\nu=0.3$. 
Then, the centered rectangular lattice is formed in the whole SC state (Fig.~\ref{fig:dia_nu=030_g=0}). 
Because of the orthorhombic symmetry of the vortex lattice, the order parameter for pairing with 
dominantly negative chirality is finite, $C_{-} \ne 0$. However, the positive chirality is favored 
by the linear coupling of magnetic field and chirality in Cooper pairs \cite{Agterberg_PRL_80_5184}, 
and therefore, $|C_{+}| \gg |C_{-}|$.  
A similar $H$-$T$ phase diagram is obtained for the isotropic case $\nu = 0$.  
Then, the triangular vortex lattice is stabilized. 

\subsection{Vortex lattice in the nematic state~($\nu>0$)}
Now we turn to the main topic of this paper. 
We study the vortex state of chiral superconductors which coexist with the nematic order. 
The GL model with a finite symmetry-breaking term ($g \ne 0$) is analyzed. 
The sign of the coupling constant $g$ is not important at all, and thus, 
we assume $g > 0$. For $g < 0$, the vortex lattice rotates $\pi/2$. 
We choose $g=0.05$ unless we explicitly state otherwise. 
The SC double transition occurs at zero magnetic field, and the splitting of 
two transition temperatures is $10\%$ of $T_{\c}^0$ for our choice of the coupling constant $g$.
This splitting is consistent with the experimental data indicating a double transition 
in URu$_2$Si$_2$ \cite{JPSJ_79_084705_Okazaki} and 
UPt$_3$ \cite{Adenwalla,Hayden,Adv_Phys_43_113_Sauls,Joynt}.

\begin{figure}[htbp]
  \centering
  \includegraphics[width=7.25cm,origin=c,keepaspectratio,clip]{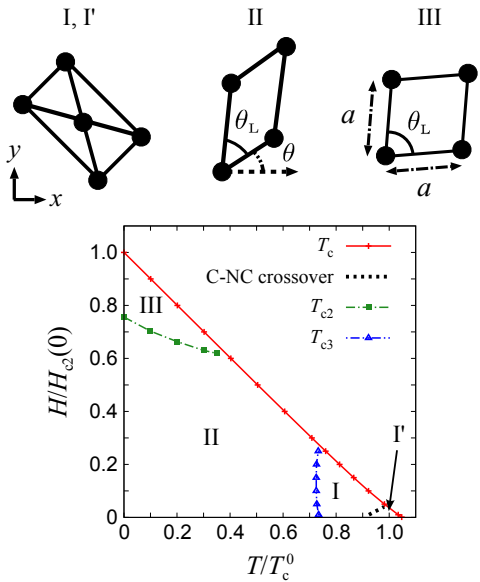}
  \caption{(Color online) Phase diagram for $(\nu,g)=(0.15,0.05)$.
    The red solid line shows the SC transition temperature $T_{\c}(H)$. 
    The green squares and blue triangles show the critical point of the vortex lattice structural transition. 
    The black dotted line depicts the C-NC crossover, which we define by $|C_{-}|=0.7|C_{+}|$.
    Phase I' is the nonchiral SC state in which the pairing function is approximately described as 
    $\De(\k) \sim \phia (\k)-\phib (\k)$, while phases I, II, and III are the chiral SC states 
    in which $\De(\k) \sim \phia (\k) + \i \phib (\k)$. 
    The vortex lattice structures in phases I (I'), II, and III are illustrated on top 
    of the phase diagram.}
  \label{fig:dia_nu=+015_g=0050}
\end{figure}

In a magnetic field $H > 0$, the chiral SC state with positive chirality $(\eta_{1}, \eta_{2}) \propto (1,0)$ 
is favored by the gradient mixing in order parameters \cite{Agterberg_PRL_80_5184}. 
Then, the second SC transition is smeared and it changes to the chiral-nonchiral crossover 
(C-NC crossover in Figs.~\ref{fig:dia_nu=+015_g=0050}, 
\ref{fig:dia_nu=+030_g=0050}, and \ref{fig:dia_nu_negative}). 
We show here that the vortex lattice structural transition occurs as a result of the C-NC crossover 
when the anisotropy parameter is positive, $\nu >0$. 
Owing to the symmetry-breaking term, the phase diagram is no longer independent of the sign of $\nu$. 
Thus, we study the case with $\nu > 0$ in this section, and the case with $\nu < 0$ is investigated 
in the next section. The isotropic case, $\nu=0$, will be discussed for UPt$_3$ in Sec.~IV.

First, we show the phase diagram for $\nu = 0.15$ in Fig.~\ref{fig:dia_nu=+015_g=0050}. 
Indeed, the vortex lattice structural transition occurs, and three phases, I (I'), II, and III, appear. 
Although the square or rectangular lattice is formed in the 
absence of the nematicity (see Fig.~\ref{fig:dia_nu=015_g=0}), a centered rectangular lattice 
is stabilized in phases I and I' near $T_{\c}$. This is intuitively understood as follows. 
Although a finite chirality is slightly induced by the gradient mixing, the order parameter 
is approximately nonchiral, $(\eta_{1}, \eta_{2}) \propto (1,-\i)$, in the high-temperature region 
near $T_{\c}$. 
Thus, an elongated triangular vortex lattice is stabilized as in single-component superconductors. 
Note that the elongated triangular lattice is equivalent to the centered rectangular lattice.

We would like to stress that the symmetry-breaking term significantly affects 
the vortex lattice structure even below the C-NC crossover temperature. 
We define the C-NC crossover line dividing phases I and I' using of the variational parameters 
as $|C_{-}| = 0.7|C_{+}|$. 
We see that the centered rectangular lattice (which equals the elongated triangular lattice) 
is stabilized well below the C-NC crossover temperature. 
Thus, the vortex lattice structure in chiral superconductors is sensitive to the nematic order. 
This is because of the small stiffness of Abrikosov vortex lattice and
the large coupling of nematicity and chirality in superconductors.

\begin{figure*}[htbp]
  \centering
  \includegraphics[width=12.25cm,origin=c,keepaspectratio,clip]{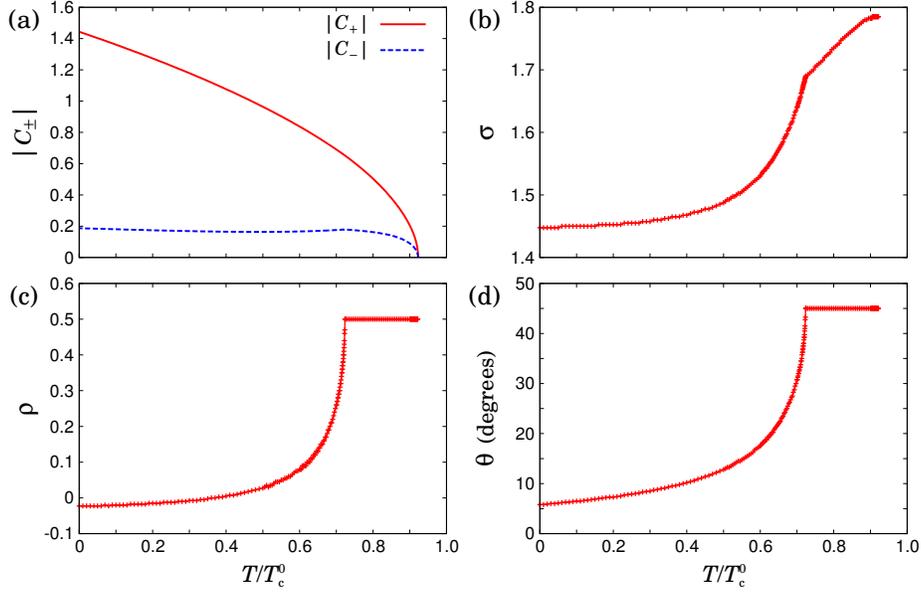}
  \caption{(Color online) Temperature dependence of (a) order parameters 
    $|C_{\pm}|$ and structural parameters (b) $\sigma$, (c) $\rho$, and (d) $\theta$ at
    $H/H_{\c 2}(0)=0.1$ for $(\nu,g)=(0.15,0.05)$.}
  \label{fig:va_tem_H=01}
\end{figure*}

\begin{figure}[htbp]
  \centering
  \includegraphics[width=6.75cm,origin=c,keepaspectratio,clip]{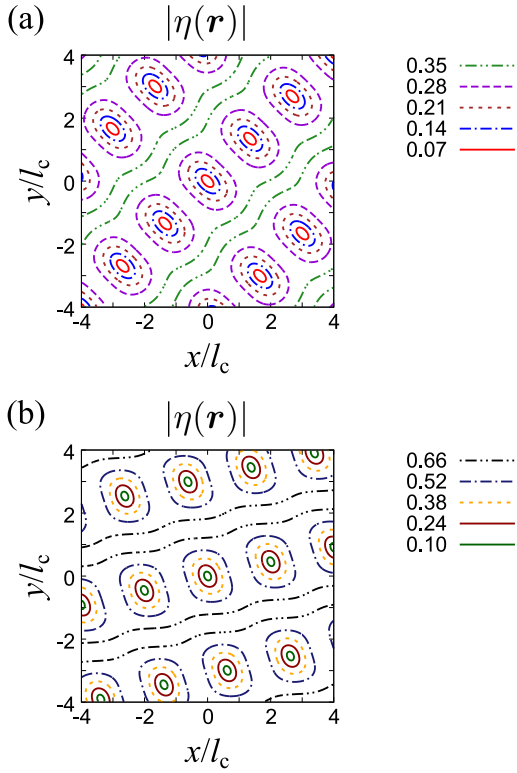}
  \caption{(Color online) 
  Vortex lattice structures (a) in phase I and (b) in phase II for $(\nu,g)=(0.15,0.05)$.
  We choose $T/T^0_{\c} = 0.8$ and $H/H_{\c 2}(0) = 0.1$ in (a) and
  $T/T^0_{\c} = 0.5$ and $H/H_{\c 2}(0) = 0.1$ in (b). 
  We show the amplitude of the order parameter $|\eta(\r)|$,
  which is obtained by $|\eta(\r)|^2 = |\eta_1(\r)|^2 + |\eta_2(\r)|^2$.
  }
  \label{fig:vortex}
\end{figure}

The vortex lattice structure in phase II is interpreted to be a distorted rectangular lattice. 
A nematicity gives rise to a stress along the [110] axis, and therefore, 
the rectangular lattice is deformed to an oblique lattice ($\rho \neq 0, 1/2$) in phase II. 
In particular, a marked effect of nematicity appears near the second-order phase transition 
to phase I, as we show the temperature dependence of variational parameters in Fig.~\ref{fig:va_tem_H=01}. 
The structural parameter is $\rho = 1/2$ in phase I (centered rectangular lattice) and decreases to 
$\rho = 0$ (rectangular lattice) with decreasing temperature in phase II. 
At the same time the angle of the primitive vector $\theta$ rotates from $\theta = \pi/4$ to $\theta \sim 0$. 
The second-order transition from phase I to phase II is accompanied by the spontaneous violation 
of the reflection symmetry along the [110] axis. 
Thus, we see marked signatures of nematic order in the vortex lattice structure 
in the low-temperature phase II. 
We show typical vortex lattice structures in phases I and II 
in Fig.~\ref{fig:vortex}(a) and \ref{fig:vortex}(b), respectively.

\begin{figure*}[htbp]
  \centering
  \includegraphics[width=12.25cm,origin=c,keepaspectratio,clip]{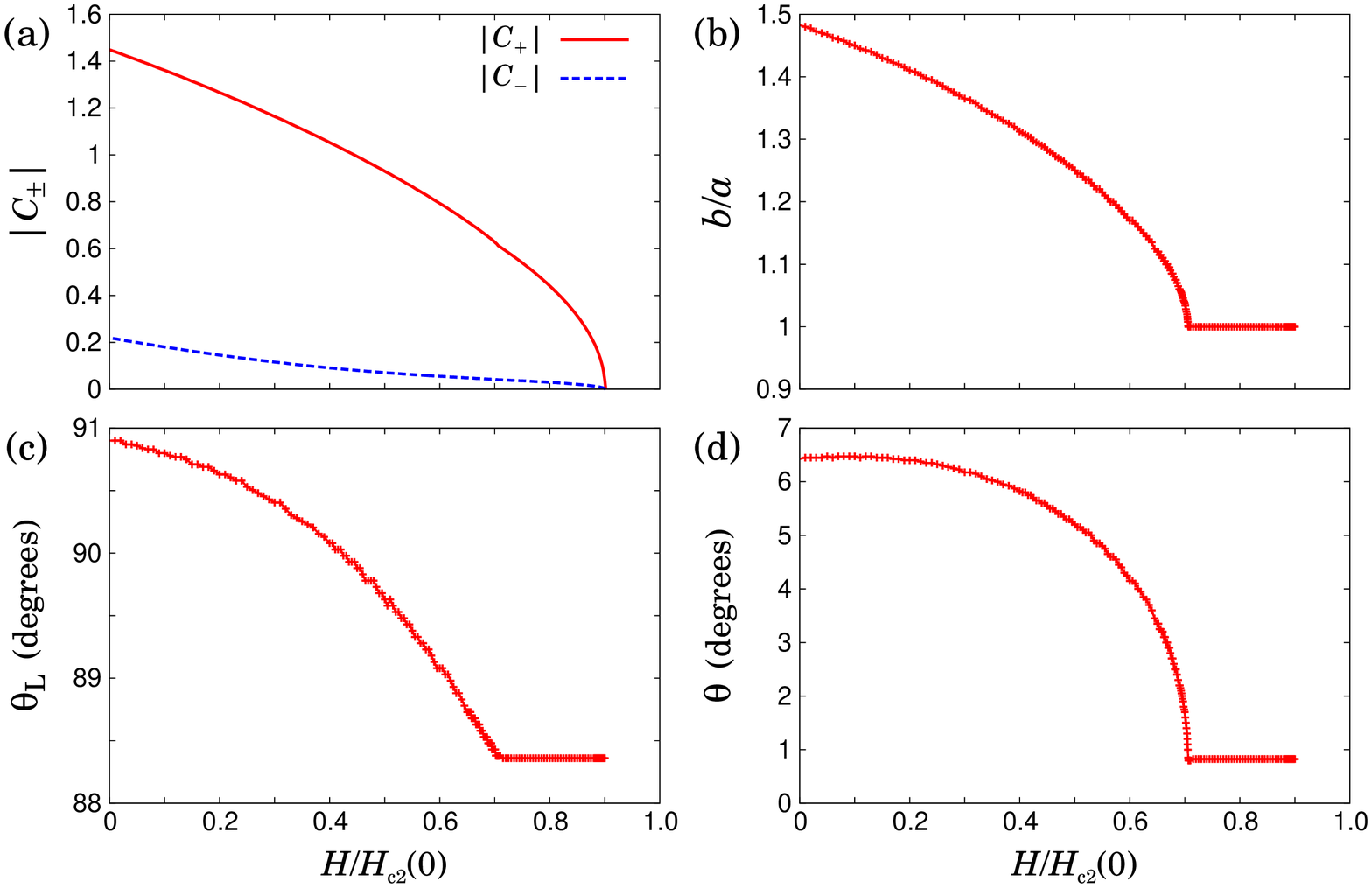}
  \caption{(Color online) Magnetic field dependence of (a) order parameters $|C_{\pm}|$ 
    and structural parameters (b) $b/a$, (c) $\theta_{\rm L}$, and (d) $\theta$ at $T/T^0_{\c}=0.1$ 
    for $(\nu,g)=(0.15,0.05)$. We show $b/a$ and $\theta_{\rm L}$ instead of $\sigma$ and $\rho$.}
  \label{fig:va_mag_t=01}
\end{figure*}

\begin{figure}[htbp]
  \centering
  \includegraphics[width=7.25cm,origin=c,keepaspectratio,clip]{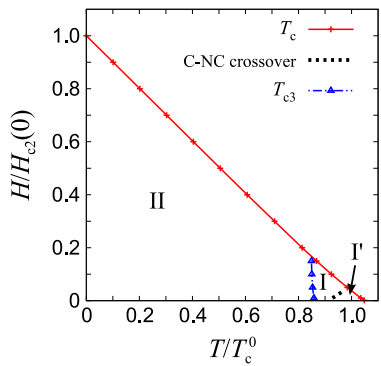}
  \caption{(Color online) Phase diagram for $(\nu,g)=(0.3,0.05)$.
    The order parameter and vortex lattice structures in phases I, I', and II are similar to those 
    in Fig.~\ref{fig:dia_nu=+015_g=0050}.} 
  \label{fig:dia_nu=+030_g=0050}
\end{figure}

The square vortex lattice is deformed to the centered rectangular lattice in the high-field phase III 
owing to the symmetry-breaking term. 
Then, $\theta_{\rm L} \ne \pi/2$; however, the deformation is negligible, as shown in Fig.~\ref{fig:va_mag_t=01}. 
The structural phase transition between phase II and phase III is characterized by the increase 
in $b/a$, as it occurs in the absence of nematicity (see Fig.~\ref{fig:nu=015_t=04}).

Although the phase transition between phases II and III is specific for the parameter $\nu =0.15$, 
the vortex lattice structural transition between phases I and II is ubiquitous as it is 
induced by the C-NC crossover. 
Indeed, the latter occurs for a wide range of the anisotropy parameter $\nu$.  
For instance, we show the phase diagram at $\nu=0.3$ (Fig.~\ref{fig:dia_nu=+030_g=0050}). 
It is shown that the vortex lattice structural transition is induced by the C-NC crossover. 
The centered rectangular lattice is stabilized in the high-temperature phases I and I', 
while the oblique lattice is stabilized in the low-temperature phase II, similar to the case of $\nu=0.15$. 
Although the centered rectangular lattice is stable in the absence of the nematicity for $\nu=0.3$
(see Fig.~\ref{fig:dia_nu=030_g=0}), it is deformed to the oblique lattice because the primitive vectors
$\bm{a}_1$ and $\bm{a}_2$ are not parallel to the stress along the [110] axis. 
The orientation of the vortex lattice is significantly rotated from $\theta = \pi/4$ to $\theta \sim 0$ 
by decreasing the temperature in phase II. 
Interestingly, a structural parameter $\rho$ shows a marked change from $\rho = 0.5$ to $\rho \sim -0.4$ 
at the same time.

Our results for $\nu =0.15$ and $\nu =0.3$ imply that the angle of a primitive vector from 
the $x$ axis $\theta$ plays an important role in the vortex lattice structural transition. 
The $x$ axis is no longer a principal axis of the electronic state for $g \ne 0$. 
Because the primitive vector should be parallel to the principal axis in the nonchiral state, 
we obtain $\theta = \pm \pi/4$ above the C-NC crossover temperature. 
The vortex lattice structural transition is induced by the C-NC crossover 
when we obtain $\theta \sim 0$ in the chiral state.  
As shown in Fig.~\ref{fig:nu_dependence}, this condition is satisfied for $\nu > 0.00114$ 
since the vortex lattice structure in the chiral state is little affected by the symmetry-breaking term. 
As expected from this consideration, the vortex lattice structural transition 
between phases I and II disappears when $\nu < -0.00114$. 
Then, $\theta= \pm \pi/4$ in the whole SC state, as we will show in the next section.

\subsection{Vortex lattice in the nematic state~($\nu<0$)}
\begin{figure}[htbp]
  \centering
  \includegraphics[width=8.5cm,origin=c,keepaspectratio,clip]{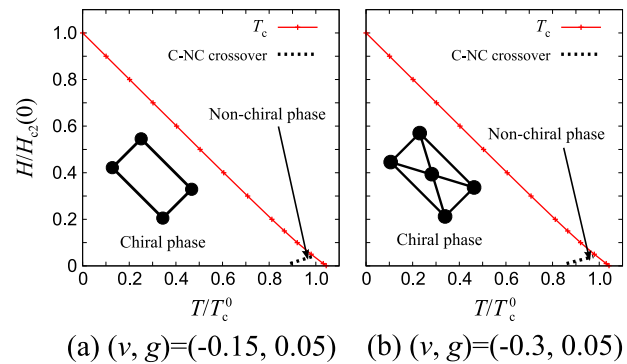}
  \caption{(Color online) Phase diagram for $\nu < 0$.
    (a) The rectangular lattice ($\rho=0, \theta=\pi/4$) is stable
    in the whole region of the SC state for $(\nu,g)=(-0.15,0.05)$, 
    while (b) the centered rectangular lattice is stable for $(\nu,g)=(-0.3,0.05)$.}
  \label{fig:dia_nu_negative}
\end{figure}

As expected from the above discussions, Fig.~\ref{fig:dia_nu_negative} shows the $H$-$T$ phase diagram 
without any structural phase transition for $\nu = -0.15$ and $\nu = -0.3$. 
The C-NC crossover occurs, as indicated by the dotted lines. 
However, the vortex lattice structural transition does not occur because the stress due to the nematicity
is applied parallel to a primitive vector of the vortex lattice ($\theta =\pi/4$). 
The rectangular lattice and centered rectangular lattice are stabilized in the whole SC state 
for $\nu=-0.15$ and for $\nu=-0.3$, respectively.
The effect of nematicity is only the increase in the parameter $\sigma$. 
The sign of the coupling constant $g$ determines the orientation of the elongated vortex lattice. 
When we choose $\sigma \geq 1$ for $\rho = 0$ and $\sigma \geq 1/2$ for $\rho = 1/2$,
the orientation is $\theta = \pi/4$ for $g>0$, while  $\theta = -\pi/4$ for $g<0$. 
For $\nu = -0.15$, the square-rectangular structural transition occurs at $g=0$,
but it is smeared by the external nematic order. The square lattice is deformed to 
the rectangular lattice owing to nematic order.

\section{Summary and Discussion}
In this paper we studied chiral SC states coexisting with a nematic order.
The phase diagram of the order parameter and vortex lattice structure has been clarified  
on the basis of the two-component GL model. 
It is shown that the vortex lattice structure is sensitive to the chirality in Cooper pairs as well as 
to the nematicity in the electronic state. Because the chirality and nematicity cooperate 
in a nontrivial way, various vortex lattice structures are stabilized. 
In particular, the structural phase transition occurs in the vortex lattice
when the nematicity is along the [110] axis and the anisotropy parameter is positive $\nu >0$, 
or when the nematicity is along the [100] axis and $\nu <0$ \cite{comment1}. 
Otherwise, the vortex lattice structural transition is not induced by the nematic order, 
although the structural parameters are affected by the C-NC crossover in the pairing function. 
The structural phase transition and structural change discussed in this work are distinguished from 
those due to the anisotropy in the Fermi surface and SC gap \cite{Wilde_PRL_78_4273,Nakai_PRL_89_237004}, 
gradient mixing in several irreducible representations \cite{Affleck_PRB_55_R704}, 
and the Pauli depairing effect \cite{Bianchi,Hiasa-Ikeda}. 
The former occurs in the low-magnetic-field region near $T_{\c}$, 
although the latter appears in the high-magnetic-field region.

Finally, we discuss the vortex state in the chiral superconductors on the basis of our results. 
Our study has been mainly focused on a heavy-fermion superconductor URu$_2$Si$_2$ \cite{Mydosh_RMP} 
in which nematic order has attracted much attention recently. 
The nematicity observed by several experiments \cite{Okazaki, Tonegawa_cycrotron, Kambe, Tonegawa_Xray} 
is along the [110]-direction. When we assume that two heavy Fermi surfaces around the $M$ point cause 
the superconductivity, the anisotropy parameter is estimated to be $\nu = 0.41$ \cite{Takamatsu_URu2Si2}
on the basis of the band structure calculation \cite{Harima} and
the Shubnikov-de Haas measurement \cite{Aoki_SdH}. 
Thus, it is reasonable to assume a positive $\nu$, although the anisotropy parameter should 
be affected by the multiband structure. 
According to our results, it is expected that the structural phase transition will occur 
in the vortex state under the $c$-axis magnetic field. 
If it were be observed by a small-angle neutron scattering (SANS) measurement \cite{Eskildsen_review}, 
for instance, clear evidence for both nematic order and chiral superconductivity would be obtained.

SANS measurements reported a square vortex lattice in Sr$_2$RuO$_4$ \cite{Riseman}. 
The orientation is $\theta = \pi/4$, implying a small negative parameter $\nu < 0$, 
and therefore, a nematicity along the [100] axis may induce the vortex lattice structural transition. 
Then, the symmetry-breaking term deforms the square lattice to the centered rectangular lattice at 
low temperatures. 
On the other hand, the nonchiral SC state with another centered rectangular vortex lattice 
is stabilized near $T_{\c}$. 
The two vortex lattices have the same symmetry; however, the structural parameters are quite different.  
For instance, our calculation for $\nu = - 0.05$ and $g=0.01$ shows that the structural parameter 
is $2\sigma \sim 1.1$ in the chiral SC state but $2\sigma \sim 3.0$ in the nonchiral SC state. 
Thus, the vortex lattice structural change accompanying the C-NC crossover may be caused 
by the uniaxial pressure along the [100] axis. Interestingly, a significant enhancement of the 
transition temperature due to the uniaxial pressure has been observed \cite{Science_344_18_Hicks}. 
However, we should discuss the spin degree of freedom in Cooper pairs because Sr$_2$RuO$_4$ 
is considered to be a spin-triplet superconductor \cite{Mackenzie2003,Maeno2012}. 
When the spin-orbit coupling in Cooper pairs is small enough to allow the rotation of the $d$ vector, 
a helical SC state, such as $\dv = \px \hat{x} - \py \hat{y}$, is stabilized in the $c$-axis magnetic field 
\cite{Takamatsu_Sr2RuO4}. 
This state is beyond the scope of this paper, but it is expected that the nematicity affects the 
vortex lattice structure through the crossover in the pairing function. 
For instance, the nematicity along the [100] axis stabilizes a planar state 
$\dv = \px \hat{x}$ or $\dv = \py \hat{y}$ near $T_{\c}$,  
while the helical state is robust at low temperatures.

The symmetry of the pairing state in UPt$_3$ is controversial
despite intensive studies for more than three decades. 
A recent polar Kerr rotation measurement found a spontaneous time-reversal symmetry-breaking, which
implies a chiral SC state $\dv(\k) = \left[\phia (\k) \pm \i \phib (\k)\right] \hat{c}$ \cite{Kapitulnik_UPt3}, 
consistent with the suppression of the upper critical field along the $c$ axis \cite{Adv_Phys_43_113_Sauls}. 
On the other hand, NMR data indicate a rotation of the $d$ vector \cite{Tou_UPt3} and implies a helical SC state,
such as $\dv(\k) = \phia (\k) \hat{b} + \phib (\k) \hat{c}$.
As for the orbital symmetry, the $E_{\rm 2u}$ and $E_{\rm 1g}$ representations have been considered as predominant 
candidates \cite{Adv_Phys_43_113_Sauls,Joynt}. On the other hand, a recent thermal conductivity measurement is 
consistent with the $E_{\rm 1u}$ orbital symmetry with a hybrid nodal structure \cite{Machida_Izawa}. 
Among these SC states, the chiral $E_{\rm 1u}$ state and chiral $E_{\rm 1g}$ state are described by the extended 
Agterberg model, and then, the square anisotropy parameter has to be zero, $\nu=0$, 
owing to the hexagonal symmetry in the crystal lattice. 
Unfortunately, this model seems to be incompatible with several experimental results. 
First, the SC double transition is smeared in the $c$-axis magnetic field, 
although it has been observed in experiments \cite{Adenwalla,Hayden,Adv_Phys_43_113_Sauls,Joynt}. 
Second, the vortex lattice structure significantly changes through the C-NC crossover, 
although only a realignment without any structural deformation has been observed 
by a SANS measurement \cite{Huxley_SANS}. 
In order to avoid these discrepancies, we have to assume parameters in the GL model 
without relying on the extended Agterberg model.  
The parameters allowed by symmetry \cite{comment2}, 
$\kappa_2 \simeq \kappa_1$ and $\kappa_{3}=\kappa_{4} \simeq 0$, may be compatible with those experiments 
and may be consistent with the nearly isotropic upper critical field in the basal plane \cite{Machida_Izawa}. 
Although these parameters are naturally obtained by the weak-coupling theory for the $E_{\rm 2u}$ state 
\cite{Adv_Phys_43_113_Sauls}, a fine-tuning of pairing functions and Fermi surfaces is required for 
the chiral $E_{\rm 1u}$ and $E_{\rm 1g}$ states.

\begin{acknowledgments}
The authors are grateful to D. F. Agterberg, D. Aoki, H. Harima, T. Kita, T. Shibauchi, and M. Yokoyama  
for fruitful discussions. 
S. T. is supported by a Japan Society for the Promotion of Science (JSPS) Research Fellowships
for Young Scientists.
This work was supported by Grants-in-Aid for Scientific Research
(KAKENHI Grants No. 25103711 and No. 24740230) from
Ministry of Education, Culture, Sports, Science and Technology (MEXT) of Japan.
\end{acknowledgments}

\bibliography{chiral_sc}

\providecommand{\noopsort}[1]{}\providecommand{\singleletter}[1]{#1}%
\begin{thebibliography}{11}%
\makeatletter
\providecommand \@ifxundefined [1]{%
 \@ifx{#1\undefined}
}%
\providecommand \@ifnum [1]{%
 \ifnum #1\expandafter \@firstoftwo
 \else \expandafter \@secondoftwo
 \fi
}%
\providecommand \@ifx [1]{%
 \ifx #1\expandafter \@firstoftwo
 \else \expandafter \@secondoftwo
 \fi
}%
\providecommand \natexlab [1]{#1}%
\providecommand \enquote  [1]{``#1''}%
\providecommand \bibnamefont  [1]{#1}%
\providecommand \bibfnamefont [1]{#1}%
\providecommand \citenamefont [1]{#1}%
\providecommand \href@noop [0]{\@secondoftwo}%
\providecommand \href [0]{\begingroup \@sanitize@url \@href}%
\providecommand \@href[1]{\@@startlink{#1}\@@href}%
\providecommand \@@href[1]{\endgroup#1\@@endlink}%
\providecommand \@sanitize@url [0]{\catcode `\\12\catcode `\$12\catcode
  `\&12\catcode `\#12\catcode `\^12\catcode `\_12\catcode `\%12\relax}%
\providecommand \@@startlink[1]{}%
\providecommand \@@endlink[0]{}%
\providecommand \url  [0]{\begingroup\@sanitize@url \@url }%
\providecommand \@url [1]{\endgroup\@href {#1}{\urlprefix }}%
\providecommand \urlprefix  [0]{URL }%
\providecommand \Eprint [0]{\href }%
\providecommand \doibase [0]{http://dx.doi.org/}%
\providecommand \selectlanguage [0]{\@gobble}%
\providecommand \bibinfo  [0]{\@secondoftwo}%
\providecommand \bibfield  [0]{\@secondoftwo}%
\providecommand \translation [1]{[#1]}%
\providecommand \BibitemOpen [0]{}%
\providecommand \bibitemStop [0]{}%
\providecommand \bibitemNoStop [0]{.\EOS\space}%
\providecommand \EOS [0]{\spacefactor3000\relax}%
\providecommand \BibitemShut  [1]{\csname bibitem#1\endcsname}%
\let\auto@bib@innerbib\@empty

\bibitem{Mackenzie2003}
A. P. Mackenzie and Y. Maeno, Rev. Mod. Phys. \textbf{75}, 657 (2003).


\bibitem{Maeno2012}
Y. Maeno, S. Kittaka, T. Nomura, S. Yonezawa, and K. Ishida,
J. Phys. Soc. Jpn. \textbf{81}, 011009 (2012). 


\bibitem{PRL_99_116402_Kasahara}
Y. Kasahara, T. Iwasawa, H. Shishido, T. Shibauchi, K. Behnia, Y. Haga, 
T. D. Matsuda, Y. Onuki, M. Sigrist, and Y. Matsuda, 
Phys. Rev. Lett. \textbf{99}, 116402 (2007).


\bibitem{PRL_100_017004_Yano}
K. Yano, T. Sakakibara, T. Tayama, M. Yokoyama, H. Amitsuka, Y. Homma, 
P. Miranovi${\rm \acute{c}}$, M. Ichioka, Y. Tsutsumi, and K. Machida, 
Phys. Rev. Lett. \textbf{100}, 017004 (2008).


\bibitem{JPSJ_Kawasaki}  
I. Kawasaki, I. Watanabe, A. Hillier, and D. Aoki, 
J. Phys. Soc. Jpn. \textbf{83}, 094720 (2014).


\bibitem{Kapitulnik_URu2Si2}
E. R. Schemm, R. E. Baumbach, P. H. Tobash, F. Ronning, E. D. Bauer,and A. Kapitulnik,
arXiv:1410.1479.


\bibitem{Yamashita_Nature_Phys}
T. Yamashita, Y. Shimoyama, Y. Haga, T. D. Matsuda, E. Yamamoto, Y. Onuki,
H. Sumiyoshi, S. Fujimoto, A. Levchenko, T. Shibauchi, and Y. Matsuda, 
Nat. Phys. \textbf{11}, 17 (2014). 


\bibitem{Luke_Sr2RuO4}
G. M. Luke, Y. Fudamoto, K. M. Kojima, M. I. Larkin, J. Merrin, B. Nachumi, Y. J. Uemura, 
Y. Maeno, Z. Q. Mao, Y. Mori, H. Nakamura, and M. Sigrist,
Nature (London) \textbf{394}, 558 (1998).

 
\bibitem{Xia_Sr2RuO4}
J. Xia, Y. Maeno, P. T. Beyersdorf, M. M. Fejer, and A. Kapitulnik, 
Phys. Rev. Lett. \textbf{97}, 167002 (2006). 


\bibitem{Kapitulnik_UPt3} 
E. R. Schemm, W. J. Gannon, C. M. Wishne, W. P. Halperin, and A. Kapitulnik, 
Science \textbf{345}, 190 (2014).  


\bibitem{Fradkin}
E. Fradkin, S. A. Kivelson, M. J. Lawler, J. P. Eisenstein, and A. P. Mackenzie, 
Annu. Rev. Condens. Matter Phys. \textbf{1}, 153 (2010). 


\bibitem{Kivelson_2003}
S. A. Kivelson, I. P. Bindloss, E. Fradkin, V. Oganesyan,
J. M. Tranquada, A. Kapitulnik, and C. Howald,
Rev. Mod. Phys. \textbf{75}, 1201 (2003). 


\bibitem{Mackenzie_Sr3Ru2O7} 
A. P. Mackenzie, J. A. N. Bruin, R. A. Borzi, A. W. Rost, and S. A. Grigera, 
Phys. C (Amsterdam, Neth.) \textbf{481}, 207 (2012). 


\bibitem{Fernandes}
R. M. Fernandes, L. H. VanBebber, S. Bhattacharya, P. Chandra, V. Keppens, D. Mandrus, 
M. A. McGuire, B. C. Sales, A. S. Sefat, and J. Schmalian, 
Phys. Rev. Lett. \textbf{105} 157003 (2010). 
%

\bibitem{Goto}      
T. Goto, R. Kurihara, K. Araki, K. Mitsumoto, M. Akatsu, Y. Nemoto, S. Tatematsu, and M. Sato
J. Phys. Soc. Jpn. \textbf{80}, 073702 (2011). 


\bibitem{Kontani}
H. Kontani and S. Onari, Phys. Rev. Lett. \textbf{104}, 157001 (2010).


\bibitem{Yanagi}
Y. Yanagi, Y. Yamakawa, N. Adachi, and Y. Ono,
J. Phys. Soc. Jpn. \textbf{79}, 123707 (2010).


\bibitem{Mydosh_RMP}
J. A. Mydosh and P. M. Oppeneer, Rev. Mod. Phys. \textbf{83}, 1301 (2011). 


\bibitem{Okazaki}
R. Okazaki, T. Shibauchi, H. J. Shi, Y. Haga, T. D. Matsuda,
E. Yamamoto, Y. Onuki, H. Ikeda, and Y. Matsuda,
Science \textbf{331}, 439 (2011). 


\bibitem{Tonegawa_cycrotron} 
S. Tonegawa, K. Hashimoto, K. Ikada, Y.-H. Lin, H. Shishido, Y. Haga, 
T. D. Matsuda, E. Yamamoto, Y. Onuki, H. Ikeda, Y. Matsuda, and T. Shibauchi, 
Phys. Rev. Lett. \textbf{109}, 036401 (2012).


\bibitem{Kambe}
S. Kambe, Y. Tokunaga, H. Sakai, T. D. Matsuda, Y. Haga, Z. Fisk, and R. E. Walstedt,
Phys. Rev. Lett. 110, 246406 (2013). 


\bibitem{Tonegawa_Xray}
S. Tonegawa, S. Kasahara, T. Fukuda, K. Sugimoto, N. Yasuda, Y. Tsuruhara, 
D. Watanabe, Y. Mizukami, Y. Haga, T. D. Matsuda, E. Yamamoto, Y. Onuki, 
H. Ikeda, Y. Matsuda, and T. Shibauchi, 
Nat. Commun. \textbf{5}, 4188 (2014). 


\bibitem{Mito}
T. Mito, M. Hattori, G. Motoyama, Y. Sakai, T. Koyama, K. Ueda,
T. Kohara, M. Yokoyama, and H. Amitsuka,
J. Phys. Soc. Jpn. \textbf{82}, 123704 (2013).


\bibitem{JPSJ_79_084705_Okazaki}
R. Okazaki, M. Shimozawa, H. Shishido, M. Konczykowski, Y. Haga, T. D. Matsuda,
E. Yamamoto, Y. $\bar{\rm O}$nuki, Y. Yanase, T. Shibauchi, and Y. Matsuda,
J. Phys. Soc. Jpn. \textbf{79}, 084705 (2010). 


\bibitem{Adenwalla} 
S. Adenwalla, S. W. Lin, Q. Z. Ran, Z. Zhao, J. B. Ketterson, J. A. Sauls, L. Taillefer, 
D. G. Hinks, M. Levy, and Bimal K. Sarma, 
Phys. Rev. Lett. \textbf{65}, 2298 (1990). 


\bibitem{Hayden}
S. M. Hayden, L. Taillefer, C. Vettier, and J. Flouquet, 
Phys. Rev. B \textbf{46}, 8675(R) (1992). 


\bibitem{Adv_Phys_43_113_Sauls}
J. A. Sauls, Adv. Phys. \textbf{43}, 113 (1994).


\bibitem{Joynt}
R. Joynt and L. Taillefer, Rev. Mod. Phys. \textbf{74}, 235 (2002). 


\bibitem{Aeppli}
G. Aeppli, E. Bucher, C. Broholm, J. K. Kjems, J. Baumann and J. Hufnagl, 
Phys. Rev. Lett. \textbf{60}, 615 (1988). 


\bibitem{Science_344_18_Hicks} 
C. W. Hicks, D. O. Brodsky, E. A. Yelland, A. S. Gibbs, J. A. N. Bruin,
M. E. Barber, S. D. Edkins, K. Nishimura, S. Yonezawa, Y. Maeno, and A. P. Mackenzie,
Science, \textbf{344}, 283 (2014). 


\bibitem{Agterberg_PRL_80_5184}
D. F. Agterberg, Phys. Rev. Lett. \textbf{80}, 5184 (1998).


\bibitem{Agterberg_PRB_58_14484}
D. F. Agterberg, Phys. Rev. B \textbf{58}, 14484 (1998).


\bibitem{Kita_PRL_83_1846}
T. Kita, Phys. Rev. Lett. \textbf{83}, 1846 (1999).


\bibitem{Rev_Mod_Phys_63_239}
M. Sigrist and K. Ueda, Rev. Mod. Phys. \textbf{63}, 239 (1991).


\bibitem{Sigrist_review_1999} 
M. Sigrist, D. Agterberg, A. Furusaki, C. Honerkamp,
K. K. Ng, T. M. Rice and M. E. Zhitomirsky,
Phys. C (Amstredam, Neth.) \textbf{317--318}, 134 (1999). 


\bibitem{Machida_Izawa}
Y. Machida, A. Itoh, Y. So, K. Izawa, Y. Haga, E. Yamamoto, N. Kimura,
Y. Onuki, Y. Tsutsumi, and K. Machida, 
Phys. Rev. Lett. \textbf{108}, 157002 (2012).


\bibitem{comment1}
The sign of $\nu$ in Eq.~(\ref{eq:nu}) is opposite to the definition 
in Refs.~\cite{Agterberg_PRL_80_5184,Agterberg_PRB_58_14484}. 


\bibitem{comment2}
This relation is satisfied within the extended Agterberg model. 
Generally, the hexagonal $D_{\rm 6h}$ point group symmetry imposes relations
$\beta_3 = 0$ and $\kappa_{3}=\kappa_{4}=(\kappa_{1}-\kappa_{2})/2$ \cite{Adv_Phys_43_113_Sauls}. 


\bibitem{SC_rev_1_207}
I. A. Luk'yanchuk and M. E. Zhitomirsky, Supercond. Rev. \textbf{1}, 207 (1995).


\bibitem{Wilde_PRL_78_4273} 
Y. De Wilde, M. Iavarone, U. Welp, V. Metlushko, A. E. Koshelev, I. Aranson,
G. W. Crabtree, and P. C. Canfield,
Phys. Rev. Lett. \textbf{78}, 4273 (1997).


\bibitem{Nakai_PRL_89_237004}
N. Nakai, P. Miranovic, M. Ichioka, and K. Machida, Phys. Rev. Lett. \textbf{89}, 237004 (2002).


\bibitem{Affleck_PRB_55_R704}
I. Affleck, M. Franz, and M. H. Sharifzadeh~Amin, Phys. Rev. B \textbf{55}, R704 (1997).


\bibitem{Bianchi}
A. D. Bianchi, M. Kenzelmann, L. DeBeer-Schmitt, J. S. White, E. M. Forgan, 
J. Mesot, M. Zolliker, J. Kohlbrecher, R. Movshovich, E. D. Bauer, 
J. L. Sarrao, Z. Fisk, C. Petrovic, and M. R. Eskildsen, 
Science, \textbf{319}, 177 (2008). 


\bibitem{Hiasa-Ikeda}
N. Hiasa and R. Ikeda, Phys. Rev. Lett. \textbf{101}, 027001 (2008). 


\bibitem{Takamatsu_URu2Si2} 
S. Takamatsu, H. Harima, D. Aoki, and Y. Yanase (unpublished). 


\bibitem{Harima}
H. Harima (private communication). 


\bibitem{Aoki_SdH}
D. Aoki, G. Knebel, I. Sheikin, E. Hassinger, L. Malone, T. D. Matsuda, and J. Flouquet, 
J. Phys. Soc. Jpn. \textbf{81} 074715 (2012). 


\bibitem{Eskildsen_review}
M. R. Eskildsen, Front. Phys. \textbf{6}, 398 (2011). 


\bibitem{Riseman}
T. M. Riseman, P. G. Kealy, E. M. Forgan, A. P. Mackenzie, L. M. Galvin, A. W. Tyler, 
S. L. Lee, C. Ager, D. McK.Paul, C. M. Aegerter, R. Cubitt, Z. Q. Mao, T. Akima, 
and Y. Maeno, Nature (London) \textbf{396}, 242 (1998); \textbf{404}, 629 (2000). 


\bibitem{Takamatsu_Sr2RuO4} 
S. Takamatsu and Y. Yanase, J. Phys. Soc. Jpn. \textbf{82}, 063706 (2013). 


\bibitem{Tou_UPt3}
H. Tou, Y. Kitaoka, K. Ishida, K. Asayama, N. Kimura, Y. Onuki, E. Yamamoto, Y. Haga, 
and K. Maezawa, Phys. Rev. Lett. \textbf{80}, 3129 (1998). 


\bibitem{Huxley_SANS}
A. Huxley, P. Rodi${\rm \grave{e}}$re, D. McK. Paul, N. van Dijk, R. Cubitt, and J. Flouquet, 
Nature (London) \textbf{406}, 160 (2000).


\end{thebibliography}%

\end{document}